\documentclass{ws-ijmpd2}
\usepackage{graphics,latexsym}
\begin{document}

\markboth{Jan Ger\v{s}l}
{Bousso entropy bound for ideal gas of massive particles}

\newcommand{\N}{\mathcal{N}}
\newcommand{\D}{{\rm d}}
\newcommand{\e}{{\rm e}}
\newcommand{\ctv}{\hspace*{\fill}$\Box$}

\title{BOUSSO ENTROPY BOUND FOR IDEAL GAS OF MASSIVE PARTICLES}

\author{Jan Ger\v{s}l}

\address{Dept. of theoretical physics and astrophysics, Masaryk University, Kotl\'{a}\v{r}sk\'{a} 2\\
611 37 Brno, Czech Republic\\
janger@physics.muni.cz}

\maketitle 

\begin{abstract}
The Bousso entropy bound is investigated for static spherically symmetric configurations of ideal gas with Bose-Einstein and Fermi-Dirac distribution function. Gas of massive particles is considered. The paper is continuation of the previous work concerning the massless case. 
Special attention is devoted to lightsheets generated by spheres. Conditions under which the Bousso bound can be violated are discussed and 
it is shown that a possible violating region cannot be arbitrarily large and that it is contained inside a sphere of unit Planck 
radius if the number of independent spin states $g_s$ is small enough. It is also shown that the central temperature must 
exceed the Planck temperature in order to get a violation of the Bousso bound for $g_s$ not too large. The situation for higher-dimensional spacetimes is also discussed and the FMW conditions are investigated. 
\end{abstract}

\keywords{Entropy bounds; static spherically symmetric spacetimes; ideal gas.}

\section{Introduction}

The idea that entropy of matter contained in a specified region of spacetime should be 
restricted by an area of a specified two dimensional spacelike surface arose in connection with black hole thermodynamics \cite{tHooft,Suss} and a lot of work have been done in this field during last decade. 
In 1999 R. Bousso formulated a hypothesis of this kind which should be valid in general spacetime in all situations where 
quantum effects are not of great importance.\cite{Bousso2} The Bousso entropy bound states the following.  
Let $(M,g)$ be a spacetime satisfying Einstein's equations and the dominant energy condition. Let $B$ be a two 
dimensional spacelike surface in $M$. Consider a null hypersurface $L$ (called lightsheet of $B$) formed by a congruence of null 
geodesics starting from $B$ orthogonally to it such that the expansion $\theta$ of the congruence is everywhere nonpositive and each geodesic 
is terminated if $\theta\rightarrow -\infty$ (caustics) , and otherwise is extended as far as possible. If we denote $S(L)$ the entropy of matter on the lightsheet and 
$A(B)$ the area of $B$, then the Bousso entropy bound states that $S(L)\leq A(B)/4$. (In all the text we use Planck units in which 
$c=1,G=1,h=2\pi ,k=1$, where $h$ is the nonreduced Planck constant and $k$ is the Boltzmann constant.) This bound was successfully 
tested in various situations.\cite{Bousso1}\cdash\cite{Lemos} Sufficient conditions for this bound were formulated in 
purely local terms by Flanagan, Marolf and Wald (FMW conditions).\cite{FMW} For review on the Bousso bound, the holographic principle and related works see Ref.~\refcite{Bousso1}. 

A stronger version of the Bousso bound was suggested in Ref.~\refcite{FMW} which allows the lightsheet to be ended not necessarily on caustics but generally on a two-dimensional surface $B^\prime$. The generalized bound then reads $S(L)\leq (A(B)-A(B^\prime))/4$. Local sufficient conditions for this bound were formulated in Ref. \refcite{BFM}. In Ref. \refcite{ales1} it was shown that the bound will be satisfied if some plausible assumptions on matter based on quantum mechanics hold.

A generalization of the Bousso bound which includes quantum effects such as Hawking radiation was suggested in Ref.~\refcite{Strom}.

The aim of this paper is to investigate the Bousso entropy bound for the case of static spherically symmetric distributions of an ideal gas with Bose-Einstein or Fermi-Dirac distribution function which consists of particles with nonzero rest mass. The case of massless particles has already been investigated in Ref.~\refcite{ja}. Special attention is devoted to the lightsheets generated by spheres where the bound reduces to the statement that entropy inside a sphere does not exceed the quarter of area of the sphere. An analysis of the entropy scaling for self-gravitating spherical matter distributions can be found also in Ref.~\refcite{Op,ales2}.

In the first part of this paper we summarize some thermodynamical quantities characterizing the ideal gas in a stationary spacetime. Then the Einstein's equations are formulated and their behavior under rescalings of parameters describing the gas is obtained. In the next section the Bousso bound for spheres is discussed and then an analysis of the general form of the bound follows based on the theorem of Flanagan, Marolf and Wald. Finally results for higher dimensional spacetimes are presented.

\section{Thermodynamic characteristics of the gas}

In this section we summarize formulas for thermodynamic quantities characterizing an ideal gas with Bose-Einstein or Fermi-Dirac statistics in a stationary spacetime. The quantities in the case of massive particles concerned here can be obtained by straightforward generalization of the statistical derivation used for the case of massless particles.\cite{ja} 

The number density of particles in one-particle phase space following from the Bose-Einstein or Fermi-Dirac statistics reads 
\begin{equation} 
\mathcal{N}=\frac{g_s}{h^3}\left({\rm e}^{(H-\mu _0)/kT_0}+z\right)^{-1},\label{dist}
\end{equation}
where $g_s$ is the number of independent spin states, $\mu _0$ and $T_0$ are constant parameters related to chemical potential and temperature and $z=-1$ for bosons or $z=1$ for fermions. Since the particles of the gas are treated as test particles moving in a mean gravitational field of the rest of the gas, $H$ is the Hamiltonian of geodesically moving particle with a mass $m$. Namely, given the timelike Killing field $\xi ^\mu$ corresponding to stationarity, the Hamiltonian reads
\begin{equation} 
H=-g_{\mu\nu}p^\mu \xi ^\nu,
\end{equation}
where $p^\mu$ is the fourmomentum of the particle. Denoting $F=\sqrt{-\xi _\mu\xi ^\mu}$ the norm of the Killing field, the minimum energy of a particle is $E_{min}=mF$. Thus, in the case of bosons we have to require 
\begin{equation}
mF\geq \mu _0 \label{nerbo}
\end{equation}
in order to obtain nonnegative density function (\ref{dist}).  

The energy-momentum tensor of the gas has the perfect fluid form 
\begin{equation}
T_{\mu\nu}=(\rho +P)u_\mu u_\nu +Pg_{\mu\nu},
\end{equation}
where $u_\mu =\xi _\mu /F$ and the energy density and pressure are given by 
\begin{eqnarray}
\rho &=&\!\!\! 4\pi\int _{m}^\infty\!\!\D \epsilon\left[\left(\epsilon ^2-m^2\right)^{\frac{3}{2}}\!\!+m^2\left(\epsilon ^2-m^2\right)^{\frac{1}{2}}\right]\N(\epsilon F)\label{rho}\\
P&=&\!\!\! \frac{4}{3}\pi\int _{m}^\infty\D \epsilon\left(\epsilon ^2-m^2\right)^{\frac{3}{2}}\N (\epsilon F),\label{P}
\end{eqnarray}
where the notation $\N (\epsilon F)$ means the function $\N$, where $\epsilon F$ is instead of $H$. Since $\N$ satisfies the collisionless Boltzmann equation, the conservation equation $\nabla ^\mu T_{\mu\nu} =0$ holds identically no matter if Einstein's equations are satisfied or not.\cite{Stew,MTW}

Given a coordinate system $x^i$ on a spacelike hypersurface $\Sigma$ and denoting $h_{ij}$ the components of the projection tensor $h_{\mu\nu}=g_{\mu\nu}+\xi _\mu\xi _\nu /F^2$ induced on $\Sigma$, for the grandcanonical potential of the gas on $\Sigma$ one obtains 
\begin{equation}
\Omega _\Sigma =-\int _\Sigma \D ^3x\sqrt{{\rm det}h_{ij}}\ FP.
\end{equation}

The entropy of the gas on $\Sigma$ then is $S_\Sigma =-\partial \Omega _\Sigma /\partial T_0$. Thus, we get 
\begin{equation}
S_\Sigma =\int _\Sigma \D ^3x\sqrt{{\rm det}h_{ij}}\ \sigma,
\label{entropy}
\end{equation}
where $\sigma =F\partial P/\partial T_0$. For the rest density of entropy $\sigma$ one obtains 
\begin{eqnarray}
\sigma &=&\!\!\! 4\pi\frac{F}{T_0} \int _{m}^\infty \!\!\!\D \epsilon
\left[\frac{4}{3}\left(\epsilon ^2-m^2\right) ^{\frac{3}{2}}\right. \nonumber \\
&&\hspace*{2cm}\left. +\left(\epsilon ^2-m^2\right)^{\frac{1}{2}}\!\left(m^2-\epsilon\frac{\mu _0}{F}\right)\right]\N(\epsilon F).\label{sigma}
\end{eqnarray}

The mean number of particles on $\Sigma$ is given by $N_\Sigma =-\partial\Omega _
\Sigma /\partial\mu _0$. This leads to the expression $N_\Sigma =\int_\Sigma\D ^3x\sqrt{{\rm det}h_{ij}}\ n$, where the rest concentration of particles $n$ is given by
\begin{equation}
n=4\pi\int _{m}^\infty\D \epsilon\left(\epsilon ^2-m^2\right)^{\frac{1}{2}}\epsilon\ \N (\epsilon F).\label{NT}
\end{equation}

The first law of thermodynamics can be recovered in the following way. Let us consider a small region of volume $V$ in the rest space of a stationary observer. Inner energy of the gas contained in such region is $U=V\rho$, 
entropy is given by $S=V\sigma$ and number of particles is $N=Vn$. The state variables $U,S,N$ as a functions of variables $V,T_0,\mu _0$ satisfy the following relation as can be verified by explicit calculation
\begin{equation}
\D U=\frac{T_0}{F}\D S-P\D V+ \frac{\mu _0}{F}\D N.
\end{equation}    
As well as in the case of massless particles \cite{ja} this suggests that the constant parameters $T_0,\mu _0$ are related to the temperature $T$ and chemical potential $\mu $ as $T=T_0/F$ and $\mu =\mu _0/F$. This agrees with the result of Tolman \cite{Tolman}, which says that a product of $T$ and $F$ is constant across a static spacetime in thermodynamic equilibrium.

\section{Einstein's equations}

In this section we formulate the Einstein's equations for spherically symmetric static configurations of ideal gas and we discuss some properties of their solutions.

We start with a metric ansatz
\begin{equation}
\D s^2=-F(r)^2\D t^2+\left(1-\frac{2M(r)}{r}\right)^{-1}\D r^2+r^2\D \Omega ^2.\label{metan}
\end{equation} 
The Einstein's equations for the unknown functions $F(r),M(r)$ then read \cite{Wald} 
\begin{eqnarray}
M^\prime &=&4\pi r^2\rho ,\label{erM}\\
F^\prime &=&F\frac{M+4\pi Pr^3}{r(r-2M)}  .\label{erF}
\end{eqnarray}
Prime denotes derivative with respect to $r$. The energy density and pressure are functions of $F$ given by the formulas (\ref{rho}) and (\ref{P}) and they depend on $r$ only through the function $F(r)$. The equation of hydrostatic equilibrium $P^\prime =-(P+\rho)F^\prime /F$, following from the equations of motion $\nabla _\mu T^{\mu\nu}=0$, is satisfied identically.

The solution of Einstein's equations is given by the parameters $m,T_0,\mu _0 ,g_s, z$ and by the initial values of functions $M(r)$ and $F(r)$. We have to take $M(0)=0$ to avoid a central singularity. New values of parameters $F(0)_{new}=\alpha F(0)_{old}$, $T_{0new}=\alpha T_{0old}$, $\mu _{0new}=\alpha \mu _{0old}$ (the rest of the parameters remains unchanged) lead to a new solution $F(r)_{new}=\alpha F(r)_{old}$, $M(r)_{new}=M(r)_{old}$, which is isometric to the old one. Thus, we can use this gauge freedom to fix the value of $F(0)$ to $F(0)=1$ without loss of generality. 

The entropy inside a sphere of radius $r_0$ is given by 
\begin{equation}
S(r_0)=4\pi\int _0^{r_0}\D r\ r^{2}\sqrt{g_{rr}}\sigma ,\label{sr}
\end{equation}
where the entropy density $\sigma$ is given by (\ref{sigma}).

Let us denote $a=\mu _0/T_0$ and choose values of parameters $m^\ast , T_0^\ast , a^\ast ,g_s^\ast ,z^\ast$. Let us denote $\rho ^\ast(r)$ the energy density as a function of $r$ corresponding to the solution of Einstein's equations given by the parameters with stars. Similar notation is used also for the other functions. The following lemma tells us how the thermodynamic quantities and the metric components transform under some parameter transformations.
\begin{lemma}
The functions corresponding to a solution of Einstein's equations given by values of parameters $m, T_0, a,g_s,z$ related to the "star parameters" as 
\begin{equation}
g_s=\beta ^4g_s^\ast ,\ \ \ \ m=\frac{\alpha }{\beta} m^\ast ,\ \ \ \ T_0=\frac{\alpha }{\beta} T_0^\ast ,\ \ \ \ \ a=a^\ast ,\ \ \ \ z=z^\ast\label{tra}
\end{equation}
are given by
\begin{eqnarray}
\rho(r;\alpha)&=&\alpha ^{4}\rho ^\ast(\alpha ^2r),\label{trarho}\\
P(r;\alpha)&=&\alpha ^{4}P^\ast(\alpha ^2r),\label{traP}\\
\sigma (r;\alpha ,\beta)&=&\beta\alpha ^{3}\sigma ^\ast(\alpha ^2r),\label{trasigma}\\
F(r;\alpha)&=& F^\ast (\alpha ^{2}r),\label{traF}\\
M(r;\alpha)&=& \alpha ^{-2}M^\ast(\alpha ^{2}r),\label{traM}\\
g_{rr}(r;\alpha)&=& g_{rr}^\ast (\alpha ^{2}r),\label{tragrr}\\
S(r;\alpha ,\beta)&=&\beta\alpha ^{-3} S^\ast(\alpha ^{2}r).\label{traS}
\end{eqnarray}
\end{lemma}
\begin{proof}
If we suppose that the functions $\rho ^\ast(r), P^\ast(r), F^\ast(r), M ^\ast(r)$ satisfy the Einstein's equations (\ref{erM}) and (\ref{erF}) then also functions $\rho(r)$, $P(r)$, $F(r)$, $M(r)$ given by (\ref{trarho}), (\ref{traP}), (\ref{traF}) and (\ref{traM}) satisfy the Einstein's equations as can be shown by explicit calculation. The transformation formula (\ref{tragrr}) for $g_{rr}$ is then obtained using (\ref{traM}) and the form of $g_{rr}$ component given by (\ref{metan}). 

Next we have to show that the rescalings (\ref{trarho}), (\ref{traP}) correspond to the transformation of parameters (\ref{tra}) and that this transformation leads to the rescalings (\ref{trasigma}) and (\ref{traS}) for entropy density and entropy inside a sphere.

In the formulas (\ref{rho}), (\ref{P}) and (\ref{sigma}) the following types of integrals occur
\begin{eqnarray}
I^1_k&=&\!\!\! \int _{m}^\infty\D \epsilon\left(\epsilon ^2-m^2\right)^{k}\N (\epsilon F)\nonumber\\
&=&m^{2k+1}\int _{1}^\infty\D q\left(q ^2-1\right)^{k}\N (mq F)\label{I1},\\
I^2_k&=&\!\!\! \int _{m}^\infty\D \epsilon\left(\epsilon ^2-m^2\right)^{k}\!\!\epsilon\ \N (\epsilon F)\nonumber\\
&=&m^{2k+2}\int _{1}^\infty\D q\left(q ^2-1\right)^{k}\!\! q\ \N (mq F),\label{I2}
\end{eqnarray}
where we supposed $m\neq 0$ in the transformation of integration variable. 
Using (\ref{dist}) we see that the parameters in the integral $I^1_k$ occur only in combinations $m/T_0, g_sm^{2k+1}, a, z$, whereas $g_sm^{2k+1}$ occurs as multiplicative factor. For the integral $I^2_k$ the multiplicative factor is $g_sm^{2k+2}$ and otherwise the combinations are the same. Both integrals depend on $r$ only through the function $F(r)$. Using (\ref{tra}) and (\ref{traF}) we obtain the following transformation properties 
\begin{eqnarray}
I^1_{\frac{3}{2}}(r;\alpha)&=&\alpha ^{4}I^{1\ast}_{\frac{3}{2}}(\alpha ^2r),\label{traI1}\\ 
I^1_{\frac{1}{2}}(r;\alpha)&=&\alpha ^{2}\beta ^2 I_{\frac{1}{2}}^{1\ast}(\alpha ^2r),\label{traI12}\\
I^2_{\frac{1}{2}}(r;\alpha)&=&\alpha ^{3}\beta I_{\frac{1}{2}}^{2\ast}(\alpha ^2 r).\label{traI2}
\end{eqnarray}
Using these transformation properties and the formulas (\ref{rho}), (\ref{P}), (\ref{sigma}) we obtain the transformation properties for energy density, pressure and entropy density (\ref{trarho}), (\ref{traP}) and (\ref{trasigma}). 

The transformation formula (\ref{traS}) for entropy inside a sphere can be obtained using (\ref{sr}), with entropy density given by (\ref{trasigma}) and the $g_{rr}$ component of metric given by (\ref{tragrr}). A transformation $\bar{r}=\alpha ^2r$ in the integral (\ref{sr}) leads directly to the result.
\end{proof}
\begin{figure}[t]
\scalebox{0.31}{\includegraphics{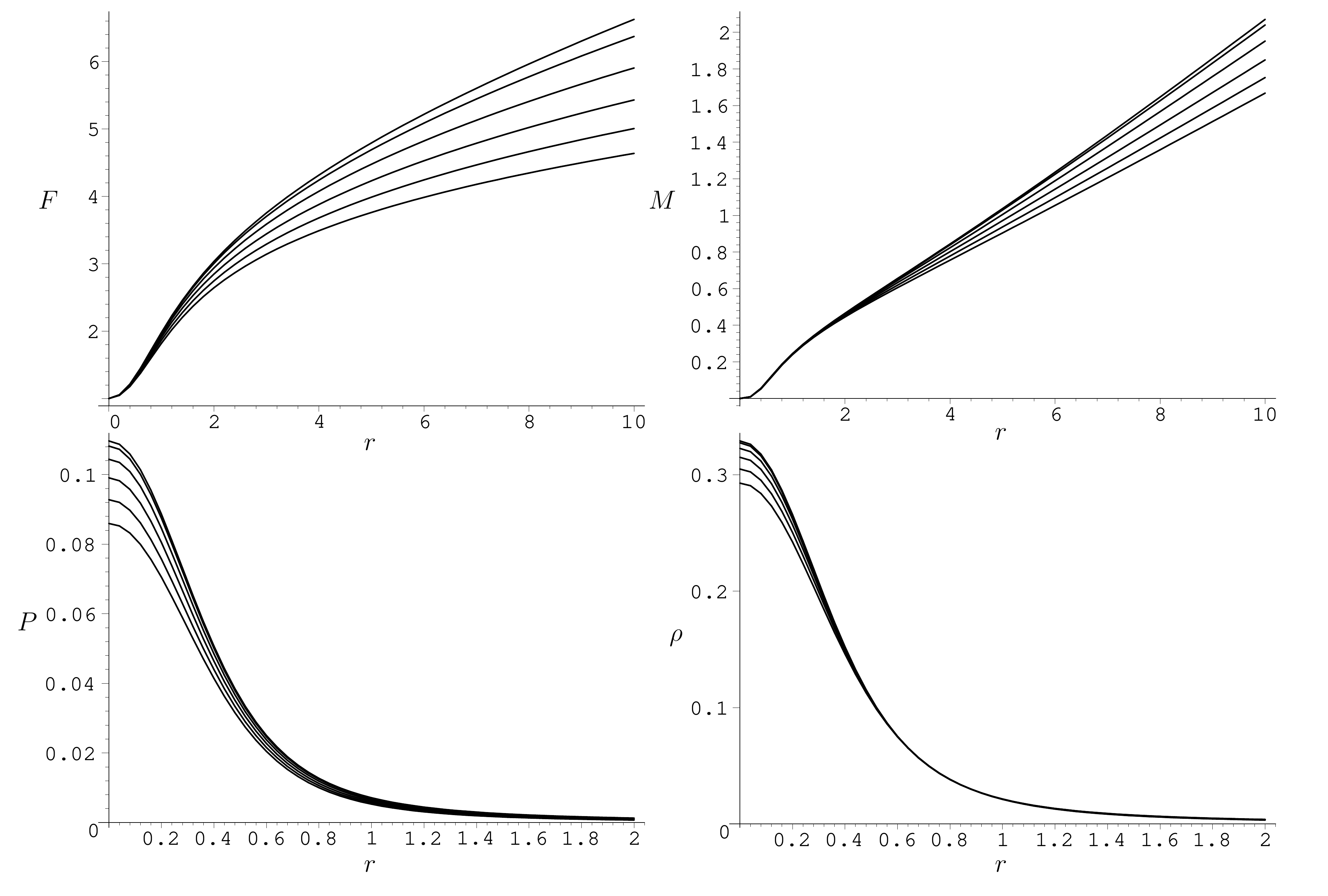}} 
\caption{The dependences $F(r),M(r),P(r),\rho(r)$ for bosons and $T_0=1,a=0,g_s=1$. From the top the curves correspond to masses $m=0,\ 0.2,\ 0.4,\ 0.6,\ 0.8,\ 1$. \label{figFm}}
\end{figure}
When solving the Einstein's equations numerically it is convenient to express the pressure and energy density in terms of special functions which are included in the function library of the program used. If we denote $y=(\epsilon F-\mu _0)/kT_0$ and we suppose that $y>0$ for all possible values of $\epsilon$, i.e. we suppose that 
\begin{equation}
mF-\mu _0>0\label{mmu}
\end{equation}
then we can write $\N(\epsilon F)$ in a form of geometric series 
\begin{equation}
\N (\epsilon F)=\frac{g_s}{h^3}(\e ^y+z)^{-1}=\frac{g_s}{h^3}\sum _{i=1}^\infty(-z)^{i-1}\e ^{-iy}.\label{gseries1}
\end{equation}
Using this expansion the integrals (\ref{I1}), (\ref{I2}) as well as the formulas for pressure, energy density and entropy density can be expressed in terms of series in modified Bessel functions of second kind \cite{wolfram} which are included in function library of MAPLE. Since bosonic gas has to satisfy the inequality (\ref{nerbo}) these series can be used to describe any ideal bosonic gas with $m\neq 0$ (the case $mF=\mu _0$ can be obtained as a limit). On the other hand for fermionic gas the inequality (\ref{nerbo}) need not be satisfied and the expansion requiring (\ref{mmu}) cannot be always used.  However, in this article we consider only the gases satisfying (\ref{nerbo}). The more complicated case of fermionic gas which does not satisfy (\ref{nerbo}) can be a matter for further research. 

Some examples of the functions $F(r), M(r), P(r)$ and $\rho(r)$ obtained with help of MAPLE are depicted in Fig.\ref{figFm}.

\section{Entropy bound}

\begin{figure}[t]
\begin{center}\scalebox{0.32}{\includegraphics{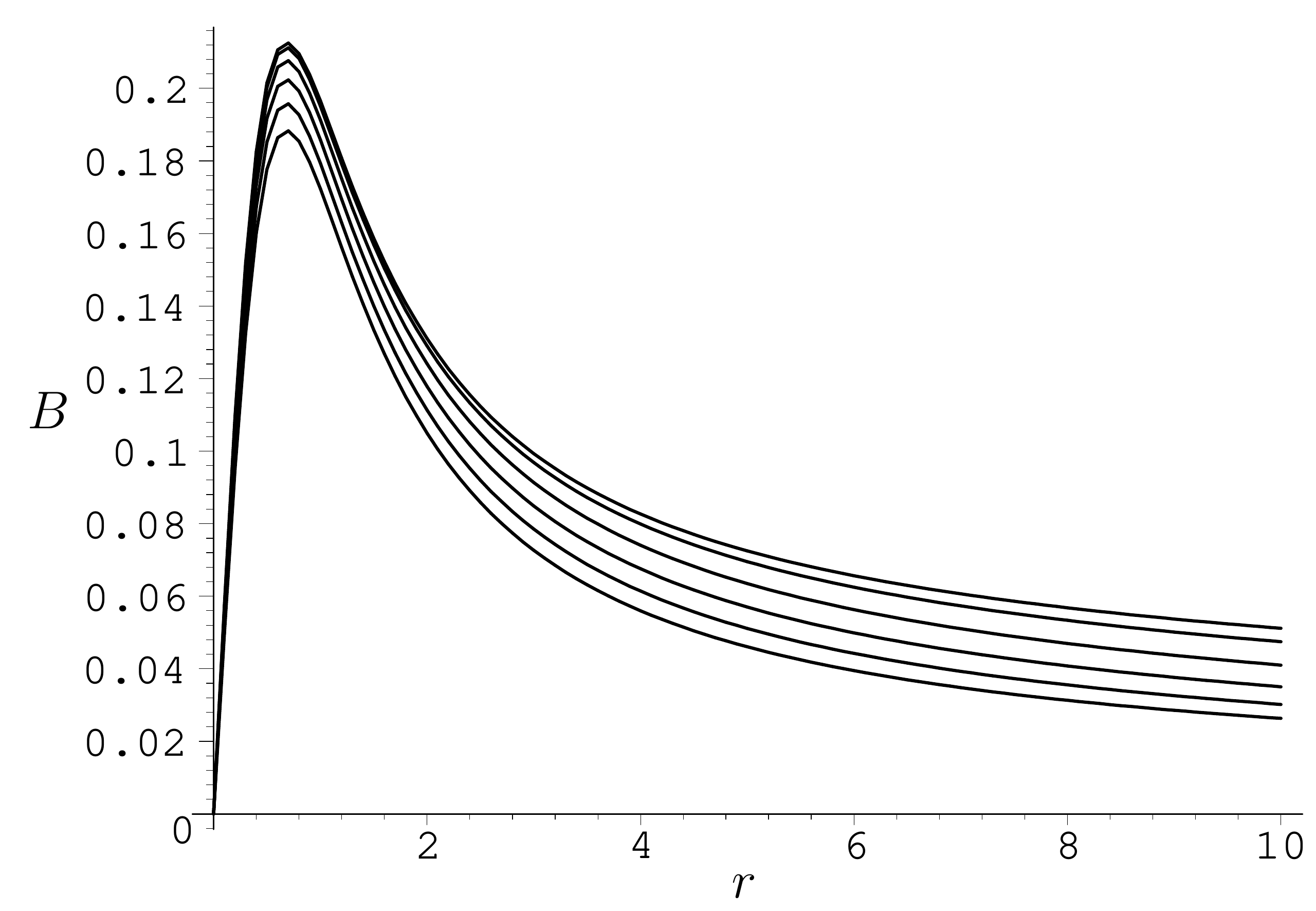}}\end{center}
\caption{The dependence $B(r)$ for bosons and $T_0=1,a=0,g_s=1$. From the top the curves correspond to masses $m=0,\ 0.2,\ 0.4,\ 0.6,\ 0.8,\ 1$. \label{figBm}}
\end{figure}

\begin{figure}[t]
\scalebox{0.23}{\includegraphics{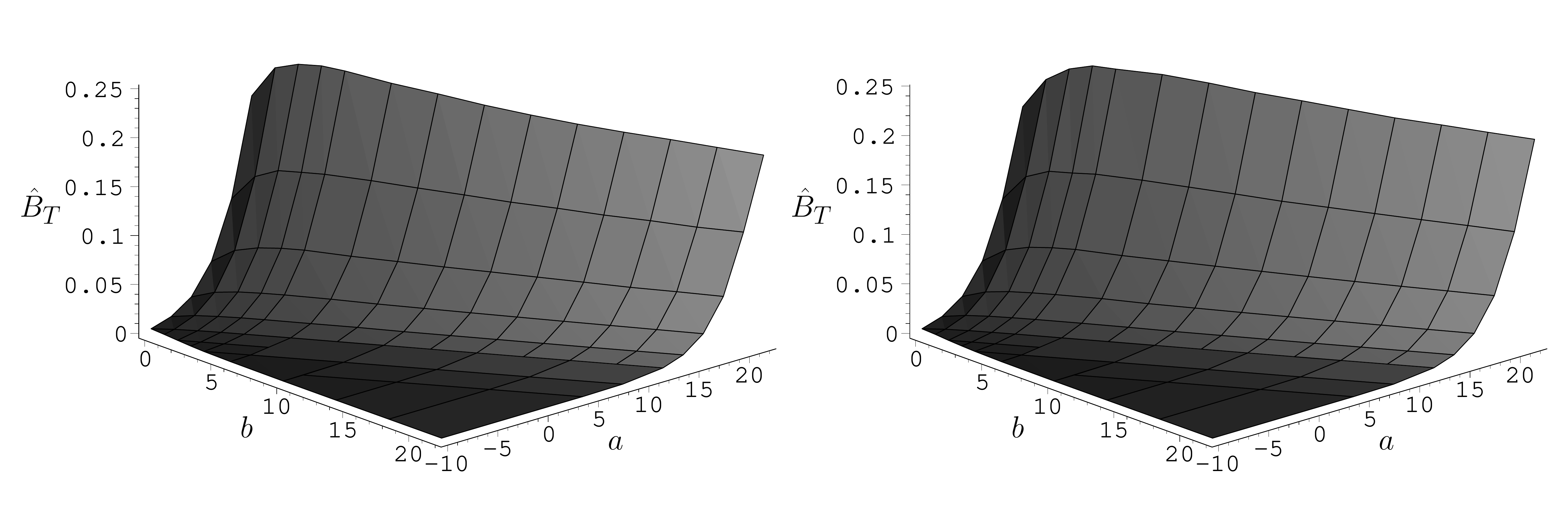}}
\caption{The dependence $\hat B_T(a,b,z)$. The picture on the left is for bosons and the second one is for fermions. \label{figd4}}
\end{figure}

\begin{figure}[t]
\begin{center}
\scalebox{0.28}{\includegraphics{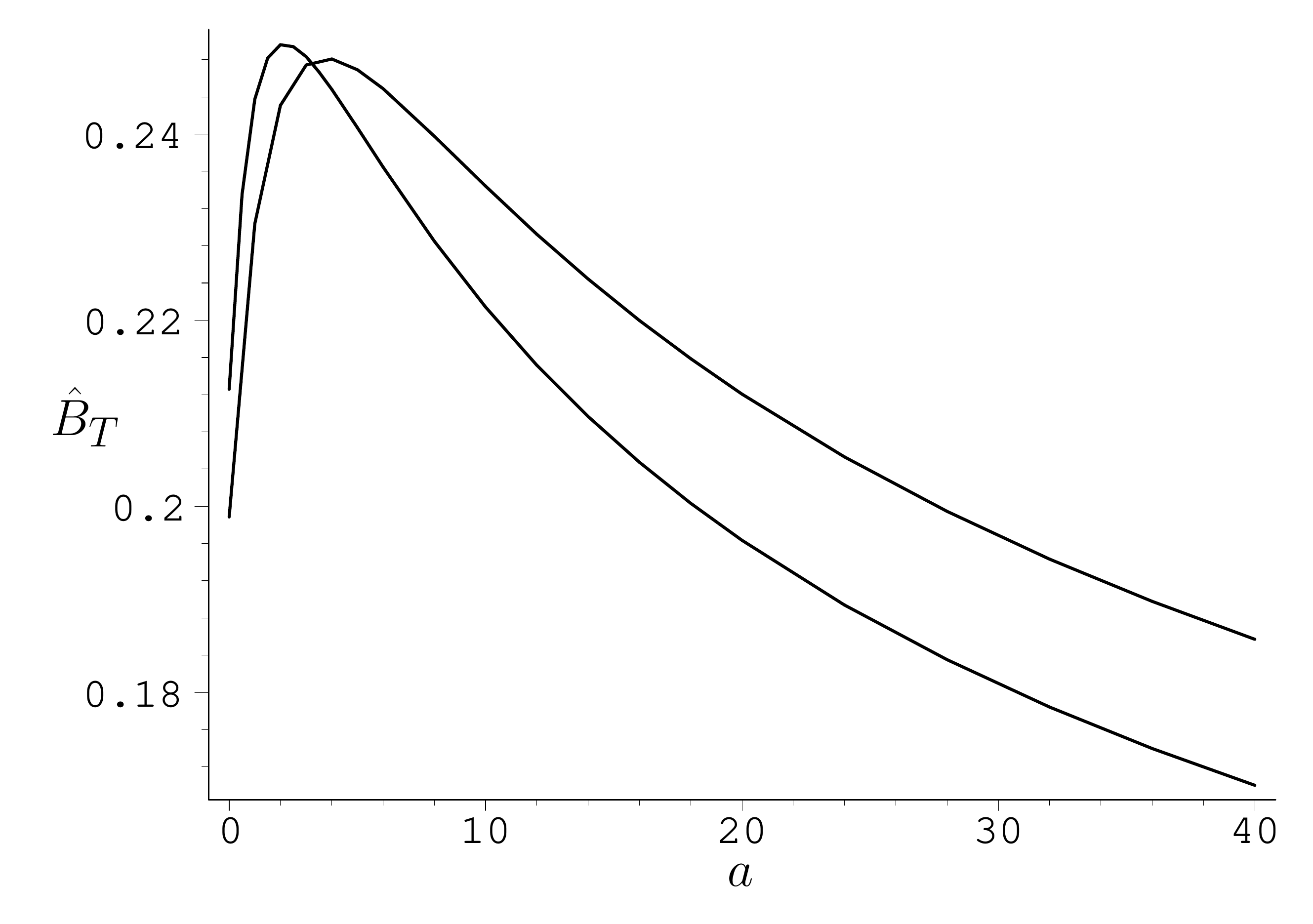}}
\end{center}
\caption{The dependence $\hat B_T(a,b,z)$ on the curve $a=b$. The curve with larger maximum is for bosons, the second one for fermions. \label{figd4top}}
\end{figure}

In this section we investigate the Bousso entropy bound for lightsheets generated by spheres. Expansion of a null geodesic 
congruence starting from a sphere orthogonally to it is $\theta =\pm 2r^{-1}EF^{-1}g_{rr}^{-1/2}$, where $+$ is for outgoing and $-$ 
for ingoing congruence and $E$ is a positive constant. Since $F>0$ and $g_{rr}>0$ everywhere, 
the lightsheet of a sphere is formed by the ingoing congruence and is ended in $r=0$. The entropy on the lightsheet generated by a sphere $r=r_0,\ t=t_0$ 
is equal to the entropy of matter inside this sphere, i.e. on the slice $r<r_0,\ t=t_0$ (see Ref.~\refcite{ja}). 
Thus, in this case, the Bousso entropy bound $S_L\leq A(B)/4$ reduces to the statement 
\begin{equation}
S(r)\leq\pi r^2
\end{equation}
in Planck units. Or, defining $B(r)=S(r)/\pi r^2$, we can write $B(r)\leq 1$. Some examples of the function $B(r)$ are depicted in Fig.\ref{figBm}. 
We see that it approaches its maximum and then it decreases to zero. Using (\ref{traS}) we obtain the following behavior under the transformation (\ref{tra})
\begin{equation}
B(r;\alpha ,\beta)=\beta\alpha B^\ast(\alpha ^{2}r).\label{traB}
\end{equation}
Les us introduce a parameter
\begin{equation}
b:=\frac{m}{T_0}
\end{equation}
and take $g_s, T_0, a, b, z$ as the full set of parameters. We use notation $B _T(r;a,b,z)$ for the function $B(r;g_s,T_0,a,b,z)$, when the values of $g_s$ and $T_0$ are fixed to $g_s=1$ and $T_0=1$. Using (\ref{traB}) and (\ref{tra}) with $g_s^\ast =1,\ T_0^\ast =1$ we obtain 
\begin{equation}
B(r;g_s,T_0,a,b,z)=g_s^{\frac{1}{2}}T_0B_T(g_s^{\frac{1}{2}}T_0^{2}r;a,b,z).\label{Bmpar}
\end{equation}
Denoting $\hat{B}_T(a,b,z)$ the maximum of $B_T(r;a,b,z)$ for given $a,b,z$, the maximum of $B(r;g_s,T_0,a,b,z)$ then reads 
\begin{equation}
B_{max}(g_s,T_0,a,b,z)=g_s^{\frac{1}{2}}T_0\hat{B}_T(a,b,z).\label{bmaxt}
\end{equation}
We see that for $T_0$ large enough $B_{max}$ can exceed unity and the Bousso bound can be violated. We can define a critical temperature $T_c$ as the temperature for which $B_{max}=1$. For $T_c$ we obtain
\begin{equation}
T_c =\left[g_s^{\frac{1}{2}}\hat{B}_T(a,b,z)\right]^{-1}.\label{tcmd}
\end{equation}
The shape of the function $\hat{B}_T(a,b,z)$ has to be found numerically. Since $m\geq 0$ and $T_0>0$ we have $b\geq 0$ and since $F(r)\geq 1$ the condition (\ref{nerbo}) can be equivalently written as $\mu _0\leq m$ or $a\leq b$. Thus, the domain of the function $\hat B_T(a,b,z)$ is given by the inequalities 
\begin{equation} 
b\geq 0,\ \ \ \ a\leq b.\label{domB}
\end{equation}  
The result of numerical calculation is depicted in Fig.\ref{figd4}. We see that for a given value of $b$ the maximum of $\hat B_T$ always occurs at $a=b$. The function $\hat B_T(a,b,z)$ restricted to the subset $a=b$ is depicted in Fig.\ref{figd4top}. 
The function $\hat B_T(a,b,z)$ has a global maximum as we can see in Fig.\ref{figd4top}. The approximate values of the maxima for bosons and fermions are
\begin{eqnarray}
{\rm max}\ \hat B_T(a,b,-1)&\approx&0.2497\\
{\rm max}\ \hat B_T(a,b,1)&\approx&0.2487.
\end{eqnarray}
The approximate values of parameters where the maxima occur are $a=b\approx 2.15$ in case of bosons and $a=b\approx 3.75$ in case of fermions. Hence, using (\ref{tcmd}), we can formulate the following lemma.
\begin{lemma}
The Bousso entropy bound for lightsheets generated by spheres in a static spherically symmetric spacetime filled with 
an ideal gas of particles with arbitrary mass satisfying $m\geq \mu _0$ is violated if and only if $T_0>T_c$, where $T_0$ and $\mu _0$ are the temperature and chemical potential of the gas in the center of symmetry and $T_c$, given by (\ref{tcmd}), satisfies
\begin{equation} 
T_c>\frac{1}{\sqrt{g_s}}({\rm max}\ \hat B_T(a,b,z))^{-1}\approx \frac{4.005}{\sqrt{g_s}}T_{Planck}.\label{tcm}
\end{equation}
\label{l2}\end{lemma}
In case of bosons the condition $m\geq \mu _0$ always holds as discussed above and if the lemma is formulated separately for bosons, the inequality for $T_c$ remains unchanged. In case of fermions we can replace ${\rm max}\ \hat B_T(a,b,z)$ by ${\rm max}\ \hat B_T(a,b,1)$. This leads to a factor 4.021 instead of 4.005 in the inequality for $T_c$. However, the analysis above does not cover all possible fermionic gases, since the condition $m\geq \mu _0$ need not be fulfilled in case of fermions.

Thus, the conclusion following from the lower bound (\ref{tcm}) is similar to the conclusion obtained in the case of massless particles.\cite{ja} If $g_s$ is of order of unity, the temperature of the gas has to exceed the Planck temperature in order to violate the Bousso bound. 

\begin{figure}[t]
\begin{center}
\scalebox{0.28}{\includegraphics{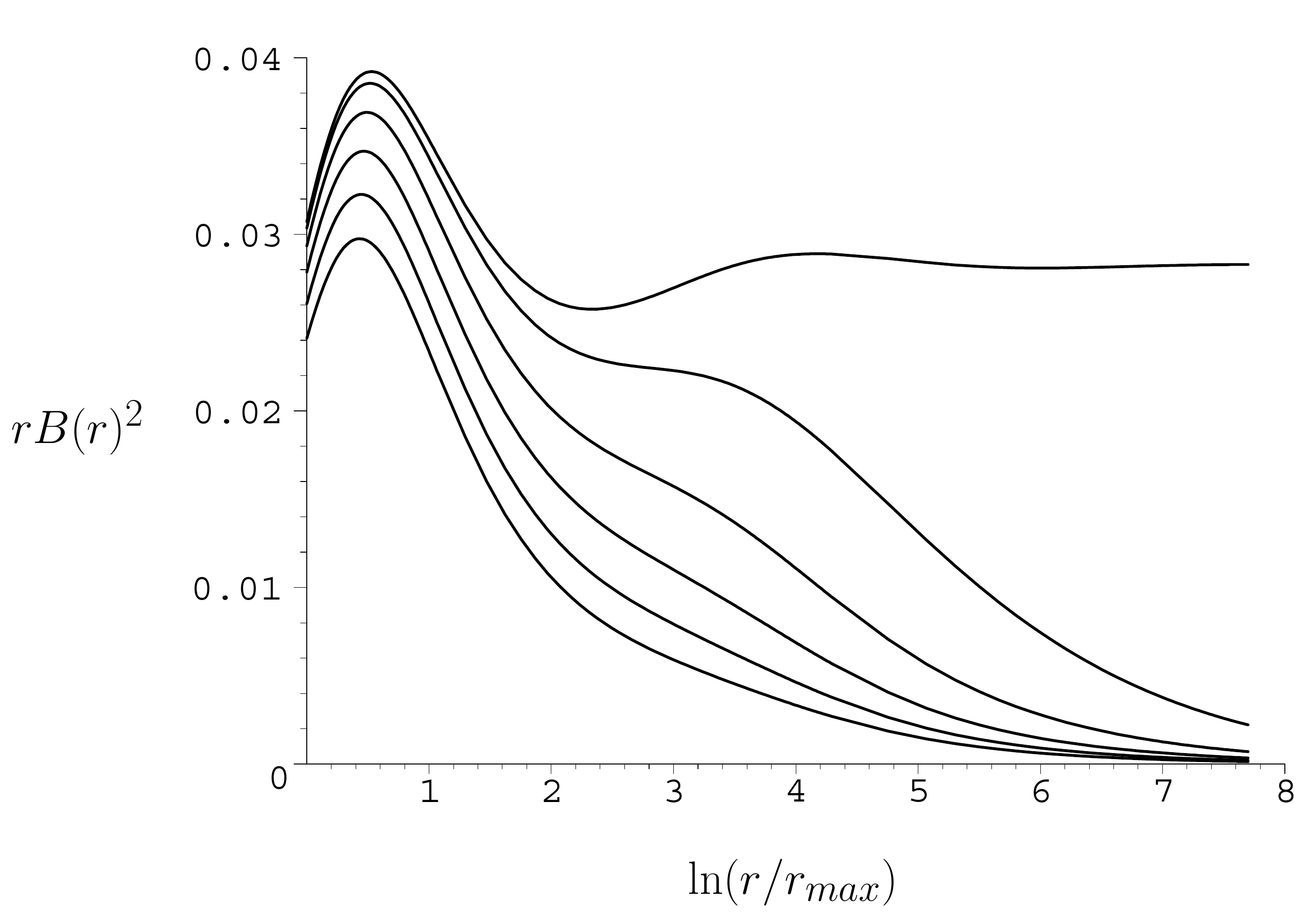}}
\end{center}
\caption{The dependence $rB(r)^2$ for $T_0=1,g_s=1,a=0,z=-1$ and various masses in logarhitmic scale. The curves correspond to masses $m=0,\ 0.2,\ 0.4,\ 0.6,\ 0.8,\ 1$ taken from the upper curve. $r_{max}$ is the value of $r$ where $B(r)$ achieves its maximum. \label{figrcm4}}
\end{figure}

When the parameters $g_s,T_0,a,b,z$ are chosen such that there is a region satisfying $B(r)>1$, the critical radius $r_c$ is defined as the larger of the two values of radial coordinate satisfying $B(r)=1$, i.e. all the spheres violating the Bousso bound are contained inside a sphere of radius $r_c$. Using (\ref{Bmpar}) and the definition formula $B(r_c;g_s,T_0,a,b,z)=1$ we obtain 
\begin{equation}
B_T(g_s^{\frac{1}{2}}T_0^{2}r_c;a,b,z)=g_s^{-\frac{1}{2}}T_0^{-1}.
\end{equation}
Denoting $C(B_T;a,b,z)$ the inverse function of the decreasing part of $B_T(r;a,b,z)$ for given values of $a,b,z$, we can express $r_c$ as 
\begin{equation} 
r_c=C(g_s^{-\frac{1}{2}}T_0^{-1};a,b,z)\ g_s^{-\frac{1}{2}}T_0^{-2}.
\end{equation}
Now we can define a parameter 
\begin{equation}
x:=C(g_s^{-\frac{1}{2}}T_0^{-1};a,b,z)
\end{equation}
instead of temperature $T_0$. In terms of the parameters $g_s,x,a,b,z$ the critical radius reads
\begin{equation}
r_c=g_s^{\frac{1}{2}}xB_T(x;a,b,z)^{2}.\label{rcdm}
\end{equation}
We are interested especially in an upper bound on $r_c$. The function $xB_T(x;a,b,z)^{2}$ for $a=0,z=-1$ and for several values of particles' mass is depicted in Fig.\ref{figrcm4}. The function $xB_T(x)^{2}$ is bounded from above for all depicted cases as well as for other values of $a,b,z$. In the case of massless particles the function approaches a nonzero value as $x$ goes to infinity.\cite{ja} From the graphs we see that adding a mass results to decrease of $xB_T(x)^{2}$ to zero as $x$ approaches infinity. Thus, in case of massive particles the value of $r_c$ goes to zero if $T_0$ approaches infinity as opposed to the massless case where the value of $r_c$ stabilizes at certain nonzero value. 

\begin{figure}[t]
\scalebox{0.25}{\includegraphics{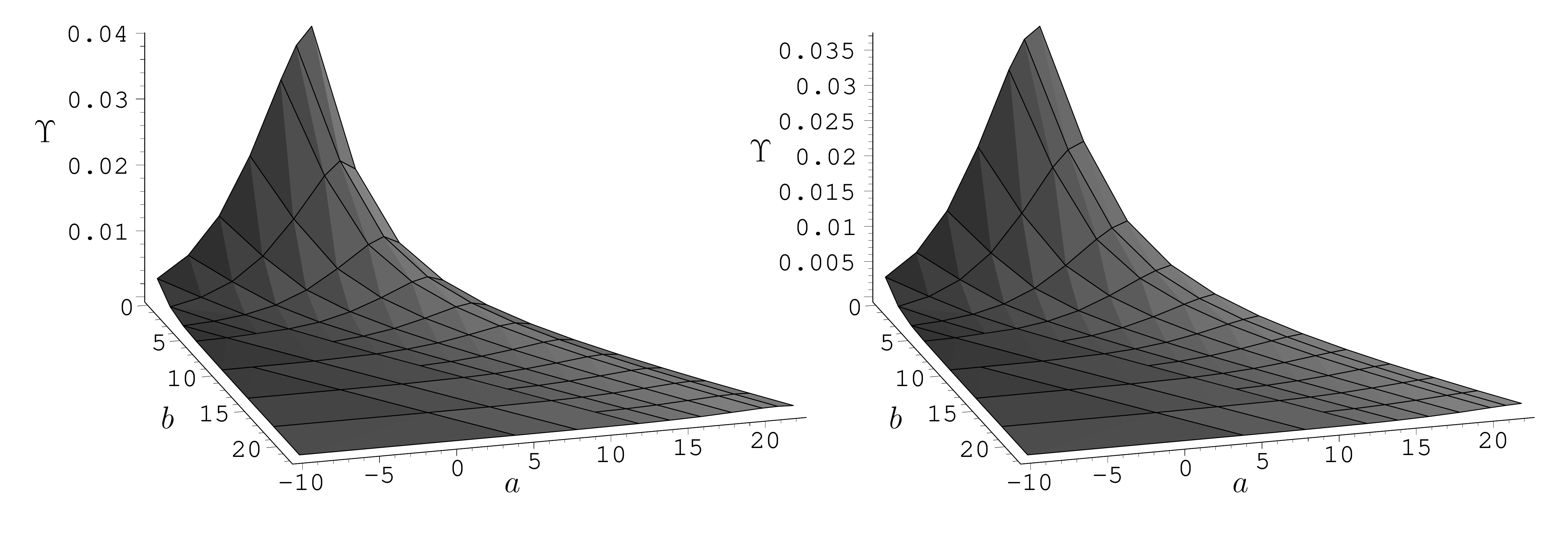}}
\caption{The dependence $\Upsilon(a,b,z)$. The picture on the left is for bosons and the second one is for fermions. \label{figy4}}
\end{figure}

The maximum of $xB_T(x)^{2}$ for given values of $a,b,z$ is denoted $\Upsilon(a,b,z)$. The domain of the function $\Upsilon(a,b,z)$ is given by inequalities (\ref{domB}) as well as the domain of $\hat B_T(a,b,z)$. The function $\Upsilon(a,b,z)$ was obtained numerically and is depicted in Fig.\ref{figy4}.
From the graphs we see that the maximum of $\Upsilon(a,b,z)$ occurs at the point $a=b=0$ for bosons and its value is $\Upsilon \approx 0.0392$ (Planck units). Since $b=0$ means $m=0$, the upper bound for $r_c$ will be identical to that for massless particles. Hence, we can formulate the following lemma.
\begin{lemma}
The Bousso entropy bound for lightsheets generated by spheres in a static spherically symmetric spacetime filled with 
an ideal gas of particles with arbitrary mass satisfying $m\geq \mu _0$ can be violated only in a region $r<r_c$, where $r_c$, given by (\ref{rcdm}), satisfies 
\begin{equation} r_c\leq g_s^{1/2} \Upsilon\approx g_s^{1/2} 0.0392 \ l_{Planck}.\end{equation}
\label{l3}\end{lemma}

\section{FMW conditions}

\begin{figure}[t]
\scalebox{0.24}{\includegraphics{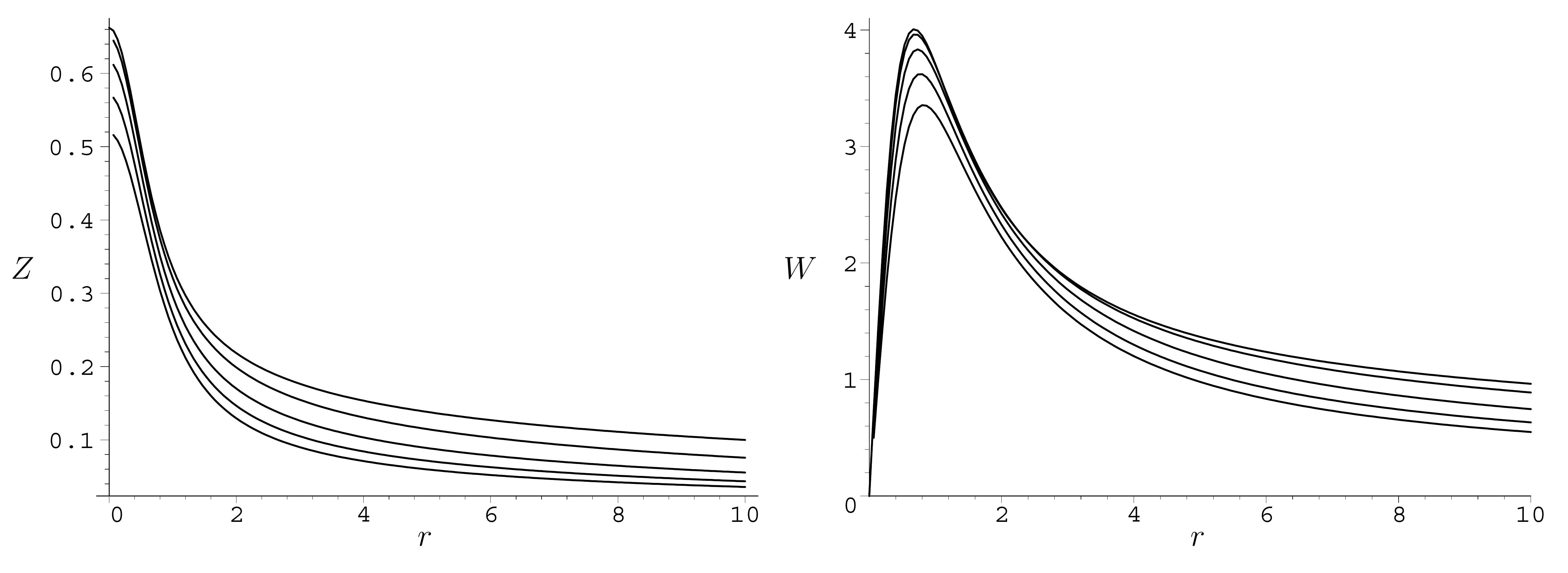}}
\caption{The dependences $Z(r)$ and $W(r)$ for bosons and $T_0=1,a=0,g_s=1$. From the top the curves correspond to masses $m=0,\ 0.5,\ 1,\ 1.5,\ 2$.  \label{figZW}}
\end{figure}

\begin{figure}[t]
\begin{center}
\scalebox{0.28}{\includegraphics{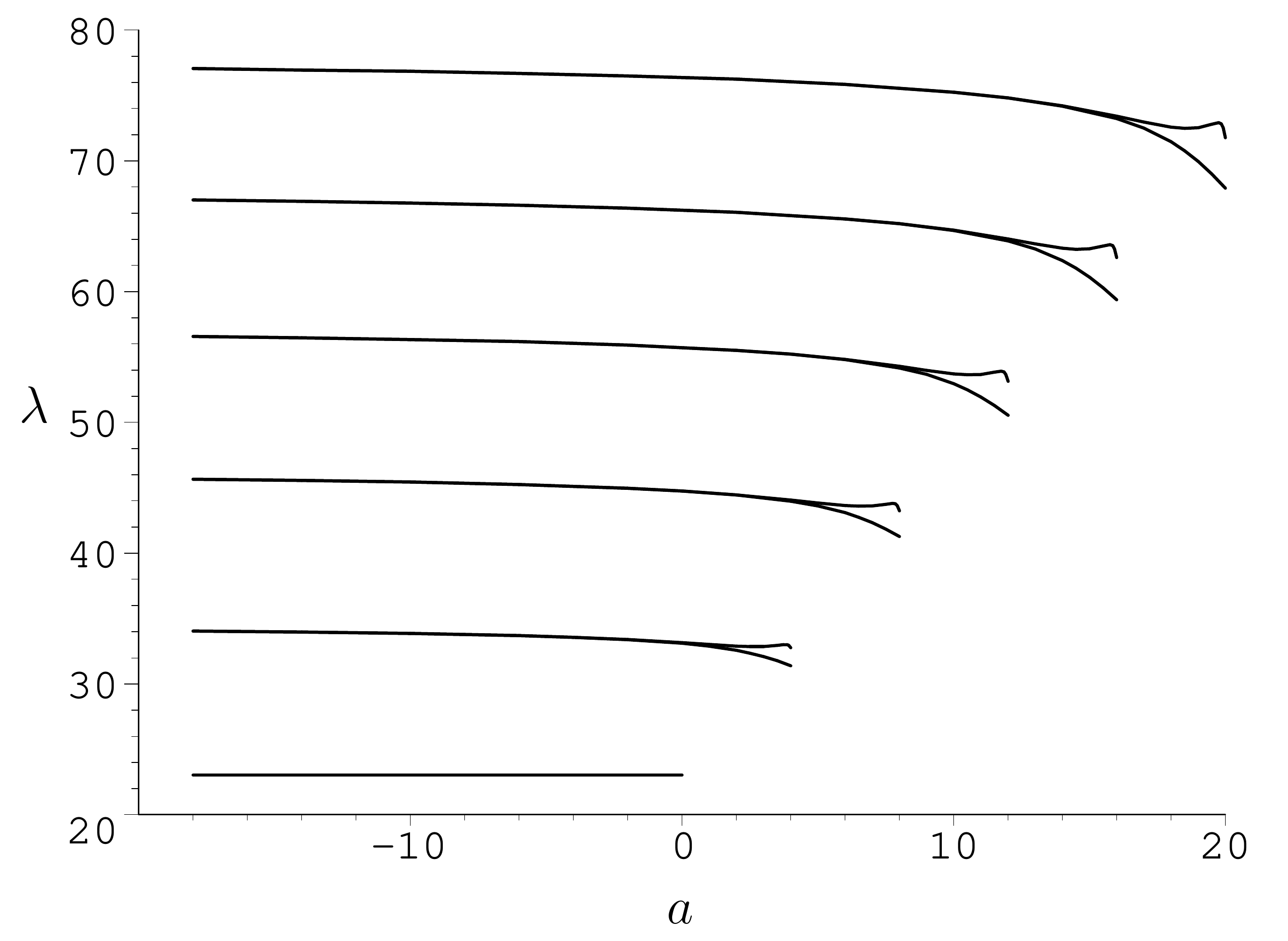}}
\end{center}
\caption{The dependence $\lambda(a,b,z)$. The upper curve from each snail tongue is for bosons and the lower one for fermions. From the top the curves correspond to $b=20,\ 16,\ 12,\ 8,\ 4,\ 0$.\label{figfmw1}}\end{figure}

The sufficient conditions for Bousso entropy bound derived by Flanagan, Marolf and Wald 
(FMW conditions) \cite{FMW} read
\begin{eqnarray} 
(s_\mu k^\mu)^2&\leq &\alpha _1T_{\mu\nu}k^\mu k^\nu, \nonumber \\
\mid k^\mu k^\nu\nabla _\mu s_\nu\mid &\leq &\alpha _2T_{\mu\nu}k^\mu k^\nu
\end{eqnarray}
for an arbitrary null vector $k^\mu$, where $s^\mu$ is an entropy flux and 
$\alpha _1,\alpha _2$ are arbitrary positive constants satisfying 
\begin{equation}
(\pi\alpha _1)^{1/4}+(\alpha _2/\pi)^{1/2}=1.\label{fm3} 
\end{equation}
In case of a stationary perfect fluid spacetime where $s^\mu=\sigma\xi ^\mu /F$ these inequalities reduce to 
\begin{eqnarray}
\sigma ^2&\leq &\alpha _1(\rho +P),\label{fm1}\\
F\sqrt{\nabla _\mu F\nabla ^\mu F} \left|\frac{\D}{\D F}\left(\frac{\sigma}{F}\right)\right|&\leq &\alpha _2(\rho +P).\label{fm2}
\end{eqnarray}
At first we will express these conditions as upper bounds for the temperature $T_0$. Defining a function
\begin{equation}
Z=\sqrt{\frac{\sigma ^2}{\rho +P}}
\end{equation}
the first condition (\ref{fm1}) can be written as $Z\leq\sqrt\alpha _1$. The function $Z(r)$ for certain values of parameters is depicted in Fig.\ref{figZW}. 
Denoting $Z_T(r;a,b,z)$ the function $Z(r;g_s,T_0,a,b,z)$ for $T_0=1$ and $g_s=1$ and using the transformation properties (\ref{trarho}), (\ref{traP}) and (\ref{trasigma}) under the transformation of parameters (\ref{tra}) with $T_0^\ast =1$ and $g_s^\ast =1$, we obtain 
\begin{equation}
Z(r;g_s,T_0,a,b,z)=g_s^{\frac{1}{2}}T_0 Z_T(g_s^{\frac{1}{2}}T_0^{2}r;a,b,z).
\end{equation}
Denoting $\hat{Z}_T(a,b,z)$ the maximum of $Z_T(r;a,b,z)$ for given $a,b,z$, the maximum of $Z(r;g_s,T_0,a,b,z)$ then reads 
\begin{equation}
Z_{max}(g_s,T_0,a,b,z)=g_s^{\frac{1}{2}}T_0\hat{Z}_T(a,b,z).\label{zmaxt}
\end{equation}
Since we require the condition $Z\leq\sqrt{\alpha _1}$ to hold in all the spacetime, it can be equivalently written as $Z_{max}\leq\sqrt{\alpha _1}$. Using (\ref{zmaxt}) the first FMW condition can be expressed as the upper bound for $T_0$
\begin{equation}
T_0\leq\alpha _1^{\frac{1}{2}}\left(g_s^{\frac{1}{2}}\hat Z_T(a,b,z)\right)^{-1}.
\end{equation}
In terms of the critical temperature (\ref{tcmd}) the first FMW condition can be expressed as
\begin{equation}
T_0\leq \alpha _1^{\frac{1}{2}}\frac{\hat B_T}{\hat Z_T}T_c.\label{fm1md}
\end{equation}
Defining a function
\begin{equation}
W=\frac{F\sqrt{\nabla ^\mu F\nabla _\mu F}\left|\frac{\D}{\D F}\left(\frac{\sigma}{F}\right)\right|}{\rho +P}
\end{equation}
the second condition (\ref{fm2}) can be written as $W\leq\alpha _2$. The function $W(r)$ for certain values of parameters is depicted in Fig.\ref{figZW}. 
In order to proceed with the "extraction" of $T_0$ from $W$ let us notice that the entropy density has a form
\begin{equation}
\sigma(r;g_s,T_0,a,b,z)=g_sT_0^3f(F(r);a,b,z).
\end{equation}
Thus, we obtain 
\begin{eqnarray}
\frac{\D}{\D F}\left(\frac{\sigma}{F}\right)(r;g_s,T_0,a,b,z) &=& g_sT_0^3\frac{\D}{\D F}\left(\frac{f(F;a,b,z)}{F}\right)\nonumber\\
&=& g_sT_0^3g(F(r);a,b,z).\label{dsdf}
\end{eqnarray}
Using (\ref{dsdf}) we obtain the same behavior under the parameter transformation (\ref{tra}) as for the entropy density $\sigma$, i.e.
\begin{equation}
\frac{\D}{\D F}\left(\frac{\sigma}{F}\right)(r;\alpha ,\beta)=\beta\alpha ^3\left[\frac{\D}{\D F}\left(\frac{\sigma}{F}\right)\right]^\ast(\alpha ^2r).\label{tradsdf}
\end{equation}
We denote $W_T(r;a,b,z)$ the function $W(r;g_s,T_0,a,b,z)$ for $T_0=1$ and $g_s=1$. Using (\ref{tradsdf}), (\ref{trarho}), (\ref{traP}), (\ref{traF}), (\ref{tragrr}) and (\ref{tra}) for $T_0^\ast =1$ and $g_s^\ast =1$ we obtain 
\begin{equation}
W(r;g_s,T_0,a,b,z)=g_s^{\frac{1}{2}}T_0 W_T(g_s^{\frac{1}{2}}T_0^{2}r;a,b,z).
\end{equation}
Denoting $\hat{W}_T(a,b,z)$ the maximum of $W_T(r;a,b,z)$ for given $a,b,z$, the maximum of $W(r;g_s,T_0,a,b,z)$ then reads 
\begin{equation}
W_{max}(g_s,T_0,a,b,z)=g_s^{\frac{1}{2}}T_0\hat{W}_T(a,b,z).\label{wmaxt}
\end{equation}
Since we require the condition $W\leq\alpha _2$ to hold in all the spacetime, it can be equivalently written as $W_{max}\leq\alpha _2$. Using (\ref{wmaxt}) the second FMW condition can be expressed as the upper bound for $T_0$
\begin{equation}
T_0\leq\alpha _2\left(g_s^{\frac{1}{2}}\hat W_T(a,b,z)\right)^{-1}.
\end{equation}
In terms of the critical temperature (\ref{tcmd}) the second FMW condition can be expressed as
\begin{equation}
T_0\leq \alpha _2\frac{\hat B_T}{\hat W_T}T_c.\label{fm2md}
\end{equation}
Denoting $\lambda ^{-1}(a,b,z)$ the smaller of the expressions standing in front of $T_c$ on the right hand sides of (\ref{fm1md}) and (\ref{fm2md}), i.e. the smaller of $\alpha _1^{1/2}\hat B_T/\hat Z_T$ and $\alpha _2\hat B_T/\hat W_T$, the FMW conditions (\ref{fm1md}) and (\ref{fm2md}) can be expressed by one inequality
\begin{equation}
T_0\leq\frac{1}{\lambda(a,b,z)}T_c.
\end{equation}
The values of $\alpha _1$ and $\alpha _2$ leading to the optimal (i.e. the largest possible) value of $\lambda ^{-1}$ are obtained as a solution of the equation 
\begin{equation}
\alpha _1^{1/2}\hat B_T/\hat Z_T = \alpha _2\hat B_T/\hat W_T
\end{equation}
together with the relation (\ref{fm3}). The resulting values of $\alpha _1,\alpha _2$ then lead to a form of $\lambda$ given by
\begin{equation}
\lambda =\pi ^{-\frac{1}{2}} \hat B_T^{-1}\left(\pi ^{-\frac{1}{4}}\hat W_T^{\frac{1}{2}}+\pi ^{\frac{1}{2}}\hat Z_T^{\frac{1}{2}}\right)^2.
\end{equation}
The function $\lambda(a,b,z)$ for certain region of parameters was obtained by numerical computation. The result is depicted in Fig.\ref{figfmw1}. 
We see that for $b=0$ (massless case) $\lambda$ is constant and its value (approximately 23.03) is the minimum of $\lambda(a,b,z)$ in the region of parameters investigated. Thus, we can conclude that: 

{\it In case of massless particles the FMW conditions led to limitation of temperature which is approximately 23 times more strict then the limitation following directly from the Bousso bound for spheres. Adding a mass leads to increase of this factor.}  

Since any maximum of $\lambda(a,b,z)$ have not been found we cannot formulate an universal bound for $T_0$ which ensures that FMW conditions will be satisfied for any values of the other parameters as it was possible for the massless case.\cite{ja}

\section{Higher spacetime dimensions}

The investigations presented above were carried out also for spacetimes of higher dimensions up to $d=11$, which should be the dimension of all including M-theory. The topology of higher dimensional spacetimes considered was ${\rm \bf R}^d$. 

\begin{figure}[t]
\begin{center}
\scalebox{0.28}{\includegraphics{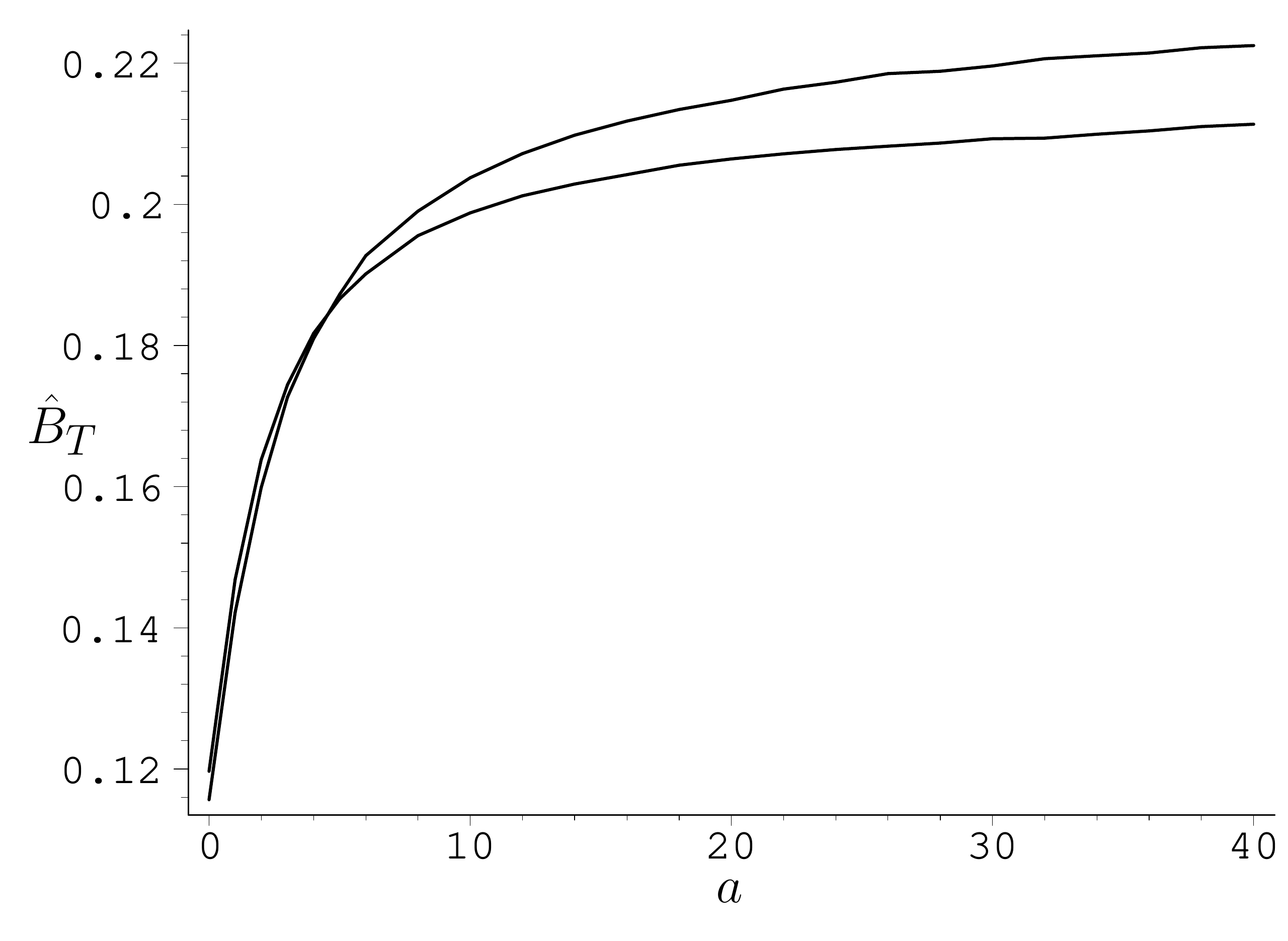}}
\end{center}
\caption{The dependence $\hat B_T(a,b,z)$ on the curve $a=b$ for $d=5$. The curve with larger values of $\hat B_T$ for large values of $a$ corresponds to fermions, the second one to bosons. \label{figd5top}}
\end{figure}

It turned out that the higher dimensional analogs of the function $\hat B_T(a,b,z)$ has no maximum (see e.g. Fig.\ref{figd5top} for $d=5$) and thus an analog of Lemma \ref{l2} cannot be formulated for the critical temperature in higher dimensions. In this sense the dimension $d=4$ is special. 

On the other hand the function $\Upsilon(a,b,z)$ has the same qualitative properties in all dimensions investigated. Especially it has maximum at $a=b=0,\ z=-1$. The values of maxima (denoted $\Upsilon$) for various dimensions are given by Tab.\ref{tabb}. Thus, the Lemma \ref{l3} can be generalized to higher dimensions as follows.
\begin{lemma}
The Bousso entropy bound for lightsheets generated by spheres in a static spherically symmetric $d$-dimensional spacetime filled with 
an ideal gas of particles with arbitrary mass satisfying $m\geq \mu _0$ can be violated only in a region $r<r_c$, where $r_c$ satisfies 
\begin{equation} r_c\leq g_s^{1/(d-2)} \Upsilon\ l_{Planck},\end{equation}
where the values of $\Upsilon$ for various dimensions are given by Tab.\ref{tabb}.
\end{lemma}
\begin{table}[ph]
\tbl{The values of the maximum of $\Upsilon(a,b,z)$ for various spacetime dimensions.}
{\begin{tabular}{lllllllll} \toprule
$d$&4&5&6&7&8&9&10&11 \\\colrule
$\Upsilon$&0.0392&0.0236&0.0169&0.0140&0.0131&0.0133&0.0146&0.0171 \\\botrule
\end{tabular} \label{tabb}}
\end{table}
Concerning the FMW conditions the function $\lambda(a,b,z)$ has always its minimum for $b=0$ (the massless case) and it is constant there. The values of this minima (denoted $\lambda$) are in Tab.\ref{tabl}. No maxima were found. Thus, the conclusions of the section on FMW conditions remain valid for the higher dimensional spacetimes up to the numerical values. 
\begin{table}[ph]
\tbl{The values of minimum of $\lambda(a,b,z)$ for various spacetime dimensions.}
{\begin{tabular}{lllllllll} \toprule
$d$&4&5&6&7&8&9&10&11 \\\colrule
$\lambda$&23.03&12.65&8.535&6.465&5.255&4.471&3.928&3.531 \\\botrule
\end{tabular} \label{tabl}}
\end{table}

\section{Conclusion}

The Bousso entropy bound in case of static spherically symmetric configurations of ideal gas of massive bosons or fermions satisfying $\mu _0\leq m$ was investigated. It was shown that for spherical lightsheets the bound can be violated only if the central temperature $T_0$ satisfies $T_0>4.005 \ g_s^{-1/2}\ T_{Planck}$. The region where the bound can be violated satisfies $r\leq \sqrt{g_s} 0.0392 \ l_{Planck}$. Similar result for the violating region was obtained also for higher dimensional spacetimes. On the other hand the lower bound on temperature in spacetimes violating the bound cannot be generalized to higher dimensions and in this sense the four-dimensional spacetime is special. 

These observations lead us to a similar conclusion as in the case of massless gas.\cite{ja} If $g_s$ is not very large which is quite plausible assumption the violation of the bound for spherical lightsheets occurs only in conditions where the classical general relativity and statistical physics are not expected to describe the physical reality. Namely in regions of the Planck scale size and for a temperature of order of Planck temperature or higher. Thus, we can say that in conditions where the physical model used is valid no violation is observed. 

Existence of a minimal scale for which statistical physics still makes sense and which protects the Bousso bound from violations was discussed e.g. in Ref.~\refcite{ales3} and it was shown that this scale should be much larger then the Planck length.

\end{document}